\documentclass[useAMS,usenatbib]{mn2e}

\usepackage{graphicx}

\title[Local analyses of Planck maps with Minkowski Functionals]
      {Local analyses of Planck maps with Minkowski Functionals}

\author[C. P. Novaes, A.  Bernui, G. A. Marques, and I. S. Ferreira]{
C. P. Novaes$^{1}$\thanks{e-mail: camilapnovaes@gmail.com} 
A. Bernui$^{1}$\thanks{e-mail: bernui@on.br} 
G. A. Marques$^{1}$\thanks{e-mail: gabrielamarques@on.br} 
and I. S. Ferreira$^{2}$\thanks{e-mail: ivan@fis.unb.br} \\
$^{1}$Observat\'orio Nacional, Rua General Jos\'e Cristino 77, 
          S\~ao Crist\'ov\~ao, 20921-400 Rio de Janeiro, RJ, Brazil \\
$^{2}$Instituto de F\'{\i}sica, Universidade de Bras\'{\i}lia, 
Campus Universit\'ario Darcy Ribeiro, Asa Norte, 70919-970, Bras\'{\i}lia, DF, Brazil 
}

\begin{document}

\date{Accepted xxxx. Received xxxx; in original form xxxx}

\pagerange{\pageref{firstpage}--\pageref{lastpage}} \pubyear{2013}

\maketitle

\label{firstpage}

\begin{abstract}
Minkowski Functionals (MF) are excellent tools to investigate the statistical properties of the cosmic 
background radiation (CMB) maps. 
Between their notorious advantages is the possibility to use them efficiently in patches of the CMB 
sphere, which allow studies in masked skies, inclusive analyses of small sky regions. 
Then, possible deviations from Gaussianity are investigated by comparison with MF obtained from a 
set of Gaussian isotropic simulated CMB maps to which are applied the same cut-sky masks. 
These analyses are sensitive enough to detect contaminations of small intensity like primary and 
secondary CMB anisotropies. 
Our methodology uses the MF, widely employed to study non-Gaussianities in CMB data, and 
asserts Gaussian deviations only when all of them points out an exceptional $\chi^2$ value, at more than 
$2.2 \, \sigma$ confidence level, in a given sky patch. 
Following this rigorous procedure, we find 13 regions in the foreground-cleaned Planck maps that 
evince such high levels of non-Gaussian deviations. 
According to our results, these non-Gaussian contributions show signatures that can be associated 
to the presence of hot or cold spots in such regions. 
Moreover, some of these non-Gaussian deviations signals suggest the presence of foreground residuals 
in those regions located near the galactic plane. 
Additionally, we confirm that most of the regions revealed in our analyses, but not all, have been 
recently reported in studies done by the Planck collaboration. 
Furthermore, we also investigate whether these non-Gaussian deviations can be possibly sourced by 
systematics, like inhomogeneous noise and beam effect in the released Planck data, 
or perhaps due to residual galactic foregrounds. 
\end{abstract}


\section{Introduction}\label{Introduction}

The cosmic microwave background (CMB) temperature fluctuations encode in their statistical properties 
unique probes of the physical processes of the early universe. 
Convincing detections of primordial non-Gaussian deviations in the CMB data, with characteristic type, 
amplitude, and scale dependence, provide invaluable information of the primeval evolution of density 
fluctuations~\citep{Bartolo04,smidt2010cmb}. 
However, non-Gaussian signals could also be sourced from late processes as, e.g., gravitational lensing, 
Sunyaev-Zel'dovich effect, contaminant radiations from galactic foregrounds 
and from residual point sources, etc., and, 
moreover, signatures from these non-Gaussian contributions could be confused with primordial ones, 
which makes necessary detailed analyses of the CMB 
data~\citep{hanson, lacasa2014, Serra, PLA2-X, PLA2-XVII}.

After accurate analyses performed on the recently released high resolution foreground-cleaned CMB maps, 
the {\it Planck collaboration} found primordial non-Gaussianity (NG) of small intensity that supports the 
simplest inflationary scenario of a single-field model \citep{PLA2-XVII}. 
In addition, complementary analyses provided strong constraints for the presence of secondary NGs in 
these maps, in particular for foregrounds residuals and contaminations from 
systematics~\citep{2015/axelsson,PLA2-XVII}.
Considering that no single statistical estimator is able to identify all possible forms of NG, 
it is worth to examine the Planck CMB data~\citep{PLA2-I} not only for confirmation of the reported results, 
but more interestingly, to perform independent analyses with diverse statistical tools applied with different 
inspection methodologies. 
For this, a variety of non-Gaussian estimators and strategies has been introduced in the literature with the 
scope to detect primary and secondary non-Gaussian contributions left in CMB data (see, for 
instance,~\cite{Babich, Bartolo10, Bartolo12, %
Noviello, Pietrobon09, Vielva04, Vielva09, 2010/matsubara, Cabella10, NN, Casaponsa11a, %
Casaponsa11b, Rossi, Smith, Donzelli, 2012/ducout, Fergusson, Pratten, BR12, BOP14, BR15, Novaes14}). 
The search for NGs in the CMB maps includes galactic and extra-galactic sources, and also possible 
contributions from systematic effects, like inhomogeneous noise, cut-sky masks, Doppler and aberration 
effects, etc.~\citep{Saha, Vielva10, Novaes15, Quartin}.

Here we use the well-known Minkowski Functionals (MF) \citep{1903/minkowski, 1999/novikov, 2003/komatsu, 2012/hikage-matsubara, 2012/munshi, 2013/munshi} to perform a detailed local analysis of the 
foreground-cleaned Planck CMB maps, 
searching for Gaussian deviations in 192 regions (i.e., pixels with size $\simeq 14.7^{\circ}$) in which 
the CMB sphere is divided. 
Our main motivations for these local non-Gaussian scrutiny are: \\
(i) to confirm previous results, that is, finding anomalous non-Gaussian regions in the CMB temperature 
field, already reported by other teams, including the Planck collaboration; \\
(ii) to identify, or at least throw some light on, the sources of the NG revealed in these anomalous regions 
by comparing their signatures imprinted in the MF with some known non-Gaussian phenomena, for 
instance, the signature left by foreground residuals or systematics (like inhomogeneous noise); \\ 
(iii) to perform complementary analyses that let us to find anomalous unreported regions. \\
In our scrutiny we found 13 regions where Gaussian deviations, as measured by the MF,
show anomalous $\chi^2$ values, at more than $2.2 \sigma$ confidence level, when compared with Gaussian 
CMB maps.

In this way, in section~\ref{sec:section2} we start presenting the Planck CMB maps used in the analysis. 
In section~\ref{sec:section3} we introduce the tools that we use to reveal NGs, that is, the MF, as well as 
the basic concepts of how to use them. 
Sections~\ref{sec:section4} and \ref{sec:section5} compiles the methodology applied to the Planck CMB 
maps and show our results, while we present our conclusions and final remarks in section~\ref{sec:section6}.

\section{The Planck data products}
\label{sec:section2}

In 2015, the Planck Collaboration released the second set of products derived from the Planck 
dataset\footnote{Based on observations obtained with Planck (http://www.esa.int/Planck), an ESA science 
mission with instruments and contributions directly funded by ESA Member States, NASA, and Canada.}. 
Between them are the four CMB foreground-cleaned maps, namely~\citep{PLA2-IX}: Spectral Matching 
Independent Component Analysis ($\mathtt{SMICA}$), Needlet Internal Linear Combination 
($\mathtt{NILC}$), Spectral Estimation Via Expectation Maximization ($\mathtt{SEVEM}$), and the 
Bayesian parametric method $\mathtt{Commander}$, names originated from the separation method used 
to produce them. 
These high resolution maps have ${\rm \,N}_{\mbox{\small side}} = 2048$ and effective beam 
\textsc{fwhm} = 5\arcmin.

The performance of the foreground-cleaning algorithms has been carefully investigated by the Planck 
collaboration with numerous sets of simulated maps \citep{PLA2-IX}. 
In particular, each component separation algorithm processed dif\/ferently the multi-contaminated sky 
regions obtaining, as a consequence, dif\/ferent final foreground-cleaned regions for each procedure. 
These regions are defined outside the so-called Component Separation Confidence masks, released 
jointly to the associated CMB map, corresponding to $f_{sky} = 0.85, 0.96, 0.84, 0.82,$ for the 
$\mathtt{SMICA}$, $\mathtt{NILC}$, $\mathtt{SEVEM}$, and $\mathtt{Commander}$ maps, respectively. 
The Planck team also released the $\mathtt{UT78}$ mask, which is the union of the above mentioned 
masks, a more restrictive one, since it has $f_{sky} = 0.776$, and adopted as the preferred mask for 
analyzing the temperature maps~\citep{PLA2-IX}. 

Our analyses were performed upon the products of the Planck's second data release. 
Furthermore, the $\mathtt{UT78}$ mask was used in all the investigated cases.

\section{The Minkowski Functionals} \label{mfs}
\label{sec:section3}
The MF present many advantages as compared with other statistical 
estimators. One of them is their versatility in detecting diverse types of NG without a 
previous knowledge of their features, such as their angular dependence or intensity, and also because 
they can be efficiently applied in masked skies or still in small regions of the CMB sphere. 

The morphological properties of data in a $d$-dimensional space can be described using $d+1$ 
MF~\citep{1903/minkowski}. 
For a 2-dimensional CMB temperature field defined on the sphere ${\cal S}^2$, 
$\Delta T = \Delta T(\theta,\phi)$, with zero mean and variance $\sigma^2$, this tool provides a 
test of non-Gaussian features by assessing the properties of connected regions in the 
map~\citep{2003/komatsu,2013/modest_mfs}. 
Given a sky path ${\cal P}$ of the pixelized CMB sphere ${\cal S}^2$, an {\em excursion set} of 
amplitude $\nu_t$ is defined as the set of pixels in ${\cal P}$ where the temperature field exceeds 
the \textit{threshold} $\nu_t$, that is, the set of pixels with coordinates $(\theta,\phi) \in {\cal P}$ such that 
$\Delta T(\theta,\phi) / \sigma \equiv \nu > \nu_t$. 
Each excursion set, or connected region, $\Sigma$, with $\nu > \nu_t$, and its boundary, 
$\partial\Sigma$, can be defined as 
\begin{eqnarray}
\Sigma & \equiv & \{(\theta, \phi) \in \mathcal{P} ~|~ \Delta T (\theta, \phi) > \nu	\sigma \}, \\
\partial\Sigma & \equiv & \{(\theta, \phi) \in \mathcal{P} ~|~ \Delta T (\theta, \phi) = \nu \sigma \}.
\end{eqnarray}
In a 2-dimensional case, for a region $\Sigma \subset {\cal S}^2$ with amplitude $\nu_t$ the 
partial MF calculated in $R_i$ are: $a_i$, the Area of the region described by 
$\Sigma$, $l_i$, the Perimeter, or contour length, $\partial\Sigma$ of this region, and 
$n_i$, the number of holes inside $\Sigma$. 
The global MF are obtained calculating these quantities for all the connected regions with height 
$\nu > \nu_t$. 
Then, the total Area $V_0(\nu)$, Perimeter $V_1(\nu)$ and Genus $V_2(\nu)$ 
are~\citep{1999/novikov, 2003/komatsu, 2006/naselsky_book, 2012/ducout}

\begin{eqnarray} \label{funcionais}
\! V_0(\nu) &=& \frac{1}{4 \pi} \int_{\Sigma} d\Omega = \sum a_i \, , \\
\! V_1(\nu) &=& \frac{1}{4 \pi} \frac{1}{4} \int_{\partial\Sigma} dl = \sum l_i \, , \\
\! V_2(\nu) &=& \frac{1}{4 \pi} \frac{1}{2 \pi} \int_{\partial\Sigma} \kappa~dl = \sum n_i = 
N_{hot} - N_{cold} \, ,
\end{eqnarray}

\noindent
where $d \Omega$ and $dl$ are, respectively, the elements of solid angle and line. 
In the Genus definition, the quantity $\kappa$ is the geodesic curvature (for details see, 
e.g.,~\cite{2012/ducout}). 
This last MF can also be calculated as the dif\/ference between the number of regions with 
$\nu > \nu_t$ (number of hot spots, $N_{hot}$) and regions with $\nu < \nu_t$ (number of cold 
spots, $N_{cold}$).

The calculations of the MF used here were done using the algorithm developed 
by~\cite{2012/ducout} and~\cite{2012/gay}.

\section{Methodology for data analyses}
\label{sec:section4}
\subsection{The local analysis} \label{sec:section4.1}

In our local analyses we use the resolution parameter ${\rm \,N}_{\mbox{\small side}} = 512$ for all CMB maps, 
data and synthetic ones, as well as for the $\mathtt{UT78}$ mask, because higher resolutions do not 
improve significantly the information contained in the MF~\citep{2012/ducout}. 
For this, we degrade the Planck CMB maps and mask to this resolution using the HEALPix (Hierarchical 
Equal Area iso-Latitude Pixelization) pixelization grid \citep{2005/gorski}. 
All pixels of the degraded mask with values different from 1 are set to 0.
After that, we define the sky patches where to carry out the local analyses
and apply the MF in each region of the CMB sky individually. 
These regions are defined by the pixels corresponding to a resolution of ${\rm \,N}_{\mbox{\small side}} = 4$, 
that is, 192 pixels of equal area $\simeq {14.7^{\circ}}^2$. 
Each of these big pixels (hereafter called {\em patches}) contains 16,384 small pixels corresponding 
to the resolution parameter ${\rm \,N}_{\mbox{\small side}} = 512$. 
We use the $\mathtt{UT78}$ mask to exclude possible galactic contaminations, consequently, 
the number of valid pixels in each region, varies from one region to another. 

Considering $n$ dif\/ferent thresholds $\nu = \nu_{_{1}}, \nu_{_{2}}, ...\, \nu_n$, previously defined 
dividing the range $- \nu_{max}$ to $\nu_{max}$ in $n$ equal parts, we compute the three MF 
$V_k, \, k=0,1,2 $, for the $p$-th patch of a CMB map, with $p = 1, 2,..., 192$. 
Then, for the $p$-th patch and for each $k$ we have the vector 
\begin{eqnarray} \label{def_v}
\,\, \mathrm{v}_k^p \equiv ( V_k(\nu_{_{1}}),V_k(\nu_{_{2}}), ... V_k(\nu_n) )|_{\mbox{\small for the $p$-th patch}} \, ,
\end{eqnarray}
for $k = 0, 1, 2$, that is, the Area, Perimeter, and Genus, respectively.
The values chosen for such variables are $(\nu_{max},n) = (3.5, 26)$ (for details, see~\cite{2012/ducout}). 

Next section presents the results of calculating the MF from Planck data and comparing them to those extracted from a set of 5000 Monte Carlo (MC) CMB maps. 
Note that such simulated data correspond to ideal CMB maps, that is, without any contribution from foreground residuals, instrumental noise or beam effect.
Our purpose here is to be able to carefully analyse the most significant differences between the ideal and real 
cases. 
These random Gaussian realisations are seeded by the CMB angular power spectrum\footnote{There is a strong 
dependence of the amplitude of the MF with the power spectrum shape \citep{2012/hikage-matsubara}. For this, we chose to use the CMB data angular power spectrum, available jointly the second Planck data release, instead of the $\Lambda$CDM best-fit. 
Thereby, the angular power spectra corresponding to synthetic and Planck CMB data are more compatible, making unnecessary to recalibrate the MF accordingly.}, 
according to the last Planck results \citep{PLA2-XI}.

\subsection{$\chi^2$ analyses of MC and Planck CMB maps}

Studies of NG in CMB data were extensively done using these three MF. 
Regarding their performances in such analyses, it was reported that the Area is the less 
sensitive to reveal NG (see, e.g.,~\cite{2012/ducout,Novaes14,Novaes15}). 
For this, we define a single 52-elements vector which combines only the Perimeter and Genus information 
into a joint estimator 
\begin{eqnarray}\label{eq7}
\mathcal{V}^p & \equiv & ( \mathrm{v}_1^p , \mathrm{v}_2^p ) \nonumber \\
                       &=& (V_1^p(\nu_1),...,V_1^p(\nu_{26}),V_2^p(\nu_1),...,V_2^p(\nu_{26}))  \, ,
\end{eqnarray}
calculated for each patch $p = 1, 2,..., 192$. 
To avoid excess of indices, in what follows we do not explicitly write the super-index $p$ when referring 
to the quantity $\mathcal{V}$, but one understands that it is being calculated for each patch $p$ of 
the CMB map (Planck or simulated data).

\begin{figure*}
\centering
 \includegraphics[width=1\textwidth]{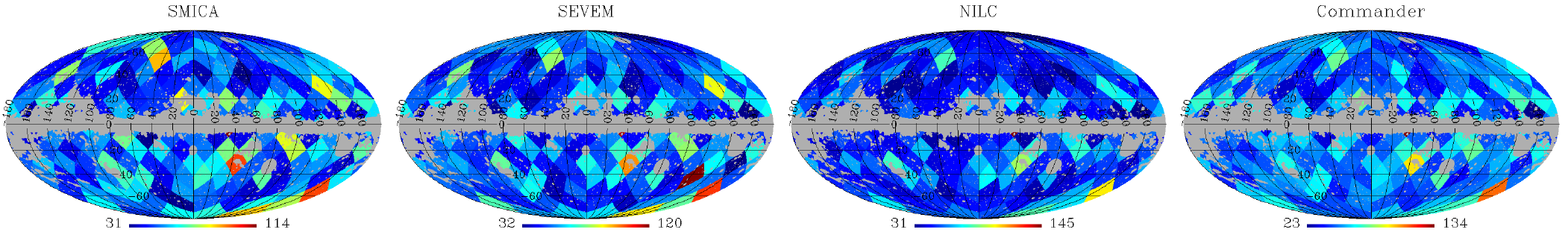}
\caption{$\chi^2$-maps obtained from the joint estimator (Perimeter and Genus) calculated from the $\mathtt{SMICA}$, 
$\mathtt{SEVEM}$, $\mathtt{NILC}$, and $\mathtt{Commander}$ Planck maps (from left to right), in 
comparison to those obtained from a set of 5000 MC CMB maps.
These $\chi^2$-maps show that the patches with the higher $\chi^2$ values are the same for all the Planck maps. 
}
 \label{fig1}
\end{figure*}

Firstly, we use the joint estimator to calculate the mean vector $\langle \mathcal{V}^{MC} \rangle$ 
for the set of 5000 MC CMB maps. 
Then, we calculate $\mathcal{V}^{Planck}$ for each Planck map to compare with the mean vector 
$\langle \mathcal{V}^{MC} \rangle$, which contains the features expected in Gaussian CMB maps. 
This is done performing a $\chi^2$ analysis which takes into account the correlations among the MF 
calculating, for each patch, the quantity~\citep{2003/komatsu,2012/ducout}
\begin{equation} \label{eq:chi2}
\chi^2 \equiv \sum_{i=1}^{52} \sum_{j=1}^{52} [ \mathcal{V}_i^{Planck} - \langle \mathcal{V}_i^{MC} \rangle ] 
\,C_{i,j}^{-1}\, [ \mathcal{V}_j^{Planck} - \langle \mathcal{V}_j^{MC} \rangle ],
\end{equation}
where the $i$ and $j$ indices run over all 52 combinations of the thresholds $\nu$ 
($26$ for the Perimeter and $26$ for the Genus), and $C_{i,j}^{-1}$ is the inverse of the full covariance 
matrix, $C_{i,j}$, which is calculated from the MC simulations as\footnote{The MF calculated at a sequence 
of thresholds $\nu$, as well as for diverse MF estimators, are correlated. In order to take such property into 
account it is important to use a full covariance matrix, $C_{i,j}$, in the data 
analyses~\citep{2006/hikage,2012/ducout}.} 
\begin{equation}
C_{i,j} \equiv \langle \,( \mathcal{V}_i^{MC} - \langle \mathcal{V}_i^{MC} \rangle )\,( \mathcal{V}_j^{MC} - \langle \mathcal{V}_j^{MC} \rangle ) \,\rangle.
\end{equation}
The maps on Figure~\ref{fig1} represent, in a color scale, the $\chi^2$ values resulting from the analysis 
of each sky region individually. 
These maps, hereafter called $\chi^2$-maps, are produced for each Planck CMB map.

The $\chi^2$-maps let us to evaluate two important issues regarding the performance of our joint estimator: 
first, the sensitivity of each MF to the local features of the data maps relative to the synthetic 
ones, and second, the influence in our results of the number of valid pixels inside a patch (i.e., those not 
excluded by the application of the $\mathtt{UT78}$ mask). 

\vspace{0.4cm}
\noindent 
\textit{\textbf{(i) Sensitivity of the MF}}: 

\noindent
The $\chi^2$-maps produced using the Equations (\ref{eq7}) and (\ref{eq:chi2}) highlight some patches of the sky, 
whose $\chi^2$ value is significantly higher than the mean. 
It shows that Perimeter and Genus, combined into a joint estimator, are able to efficiently reveal 
local Gaussian deviations in the Planck maps compared to MC Gaussian CMB data. 
For the sake of completeness, we have also tested the behavior of the joint estimator in the case that 
one includes the Area in its definition, Equation (\ref{eq7}). 
In this case, the resulting $\chi^2$-maps (one for each Planck map) reveal a rather featureless map 
with an almost uniform distribution of values through the sky, evidencing a notable disagreement with the 
results presented in Figure \ref{fig1} (inclusive, the well-known anomalous Cold Spot~\citep{Vielva04} passes 
undetected). 
These outcomes confirm the lower sensitivity of the Area relative to the others, and validates 
our previous choice of a Perimeter plus Genus joint estimator.

\begin{figure}
\centering
\includegraphics[width=0.48\textwidth]{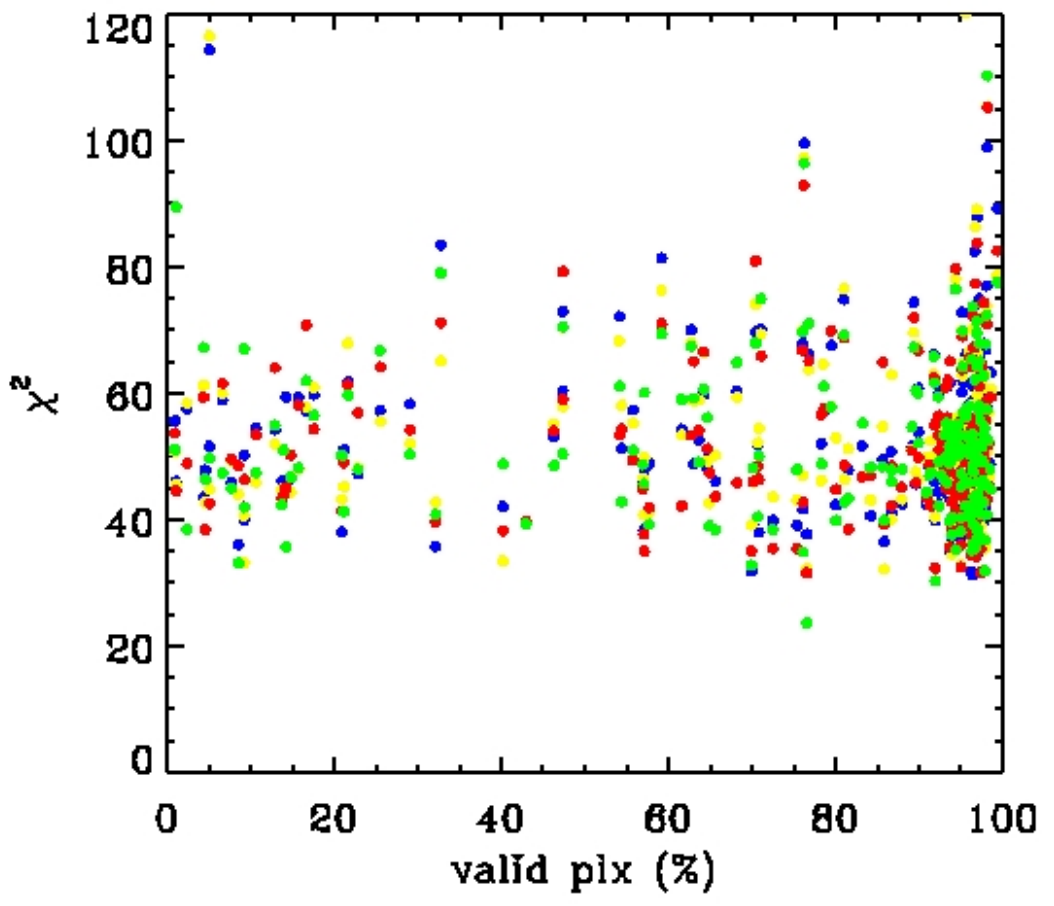}
\caption{Dependence of the $\chi^2$ values with the percentage of valid pixels 
in the patches.
The blue, yellow, red and green dots corresponds, respectively, to the results from analysing the 
$\mathtt{SMICA}$, $\mathtt{SEVEM}$, $\mathtt{NILC}$, and $\mathtt{Commander}$ 
Planck maps ($\chi^2$-maps from left to right in Figure \ref{fig1}). 
From now on, we consider patches with a minimum of $30\%$ of valid pixels. 
}
\label{fig2}
\end{figure}

\vspace{0.2cm}
\noindent
\textit{\textbf{(ii) Influence of the percentage of valid pixels}}: 

\noindent
Since all the current analysis are performed considering a sky cut given by the $\mathtt{UT78}$ mask, 
each patch of the sky is composed by a different number of valid pixels.
In this sense, it is important to check if the percentage of valid pixels influences our analysis. 
Figure \ref{fig2} shows, for each Planck CMB map, 
the dependence between the $\chi^2$ value and the percentage of valid pixels in each patch. 
This comparison allows us to observe that there is no evident relationship between these two quantities. 
Nevertheless, from now on, we consider patches with a minimum of $30\%$ of valid pixels. 

The results obtained in this section using the $\chi^2$-maps, suggest a more careful study of the set of 
patches that exhibit anomalously large $\chi^2$ values. 
For this we continue our analysis examining in detail the 
Perimeter and Genus curves corresponding to these regions, 
in order to evidence its (possible) peculiar signature that would lead to recognize its source. 
Such analyses are presented through the next section.

\section{Analysis of the anomalous regions} \label{sec:section5}

The aim of this section is the detailed analysis of the sky patches of the Planck CMB maps upon 
where the MF revealed more significant differences relatively to the MC Gaussian simulations. 
We constructed a $total \, \chi^2$-map by averaging among the $\chi^2$-maps obtained 
for each Planck CMB map (Figure \ref{fig1}), namely, the $\mathtt{SMICA}$, $\mathtt{SEVEM}$, 
$\mathtt{NILC}$, and $\mathtt{Commander}$ 

\begin{equation} \label{eq:total_chi}
total ~\chi^2 = ( \chi^2_{\mathtt{SMICA}} + \chi^2_{\mathtt{SEVEM}} + \chi^2_{\mathtt{NILC}} + \chi^2_{\mathtt{Commander}} ) / 4  \, .
\end{equation}

\begin{figure} 
\centering
 \includegraphics[width=0.49\textwidth]{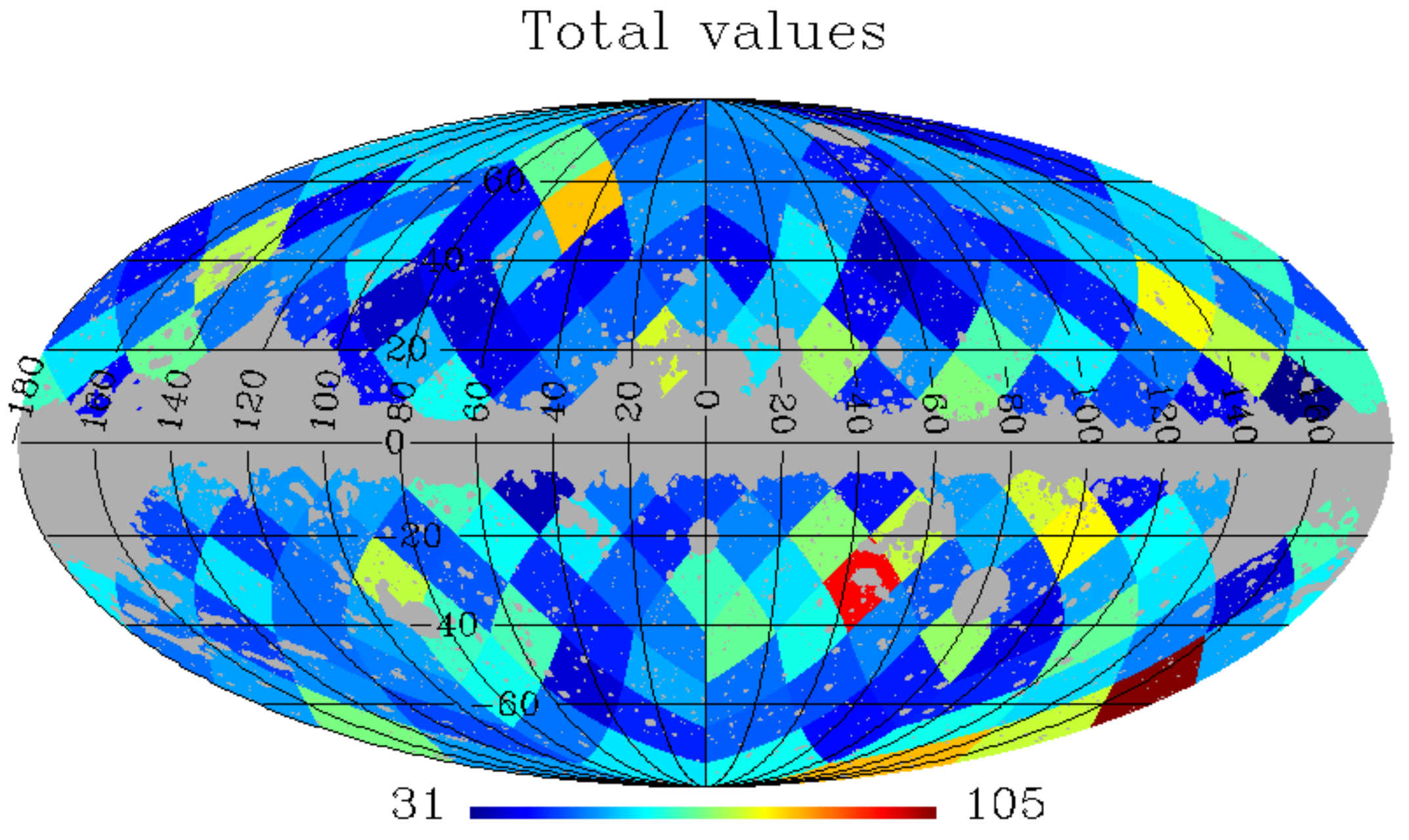}
\caption{The $total \, \chi^2$-map obtained combining the results from all the four Planck CMB maps, as given by the Equation \ref{eq:total_chi}.} 
\label{fig3}
\end{figure}

\begin{figure}
\centering
\includegraphics[width=0.48\textwidth]{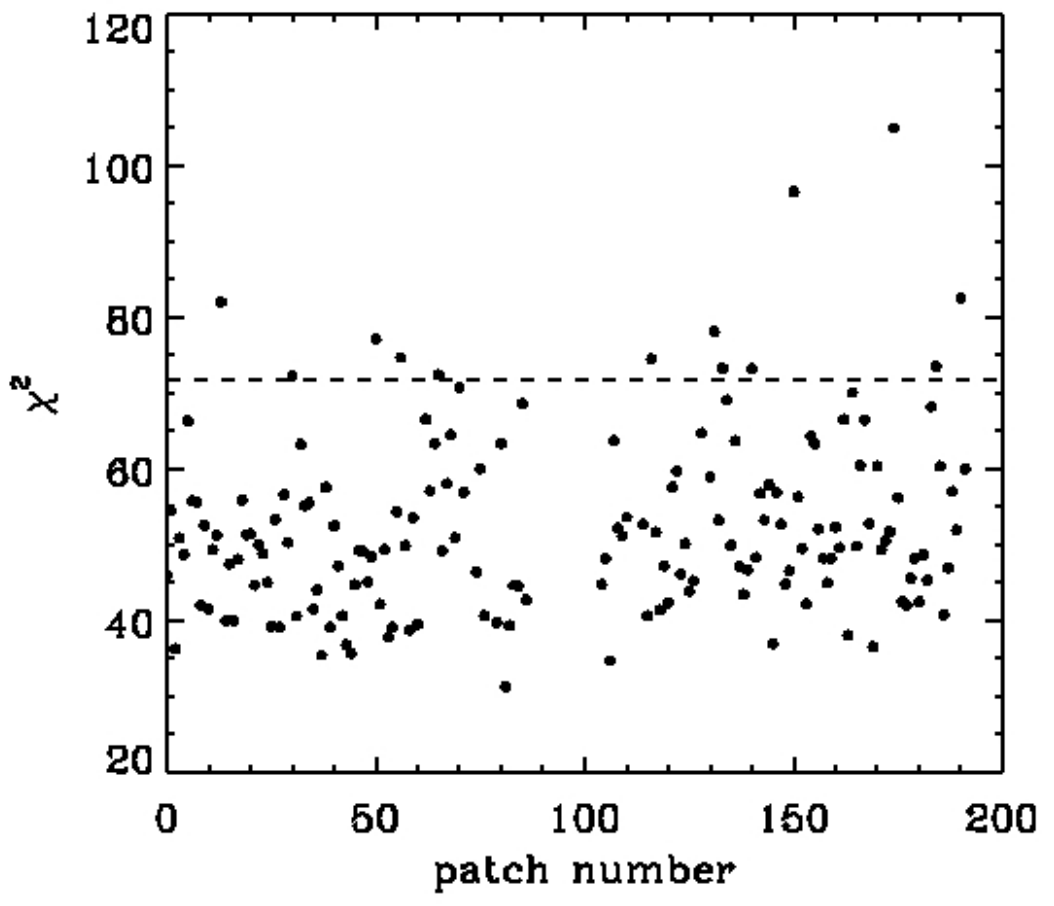}
\caption{Graphic representation of the total $\chi^2$ values as function of the patch number. 
The dashed line corresponds to the $2.2 \sigma$ value (see text for details).}
\label{fig4}
\end{figure}

\begin{figure} 
\centering
\includegraphics[width=0.49\textwidth]{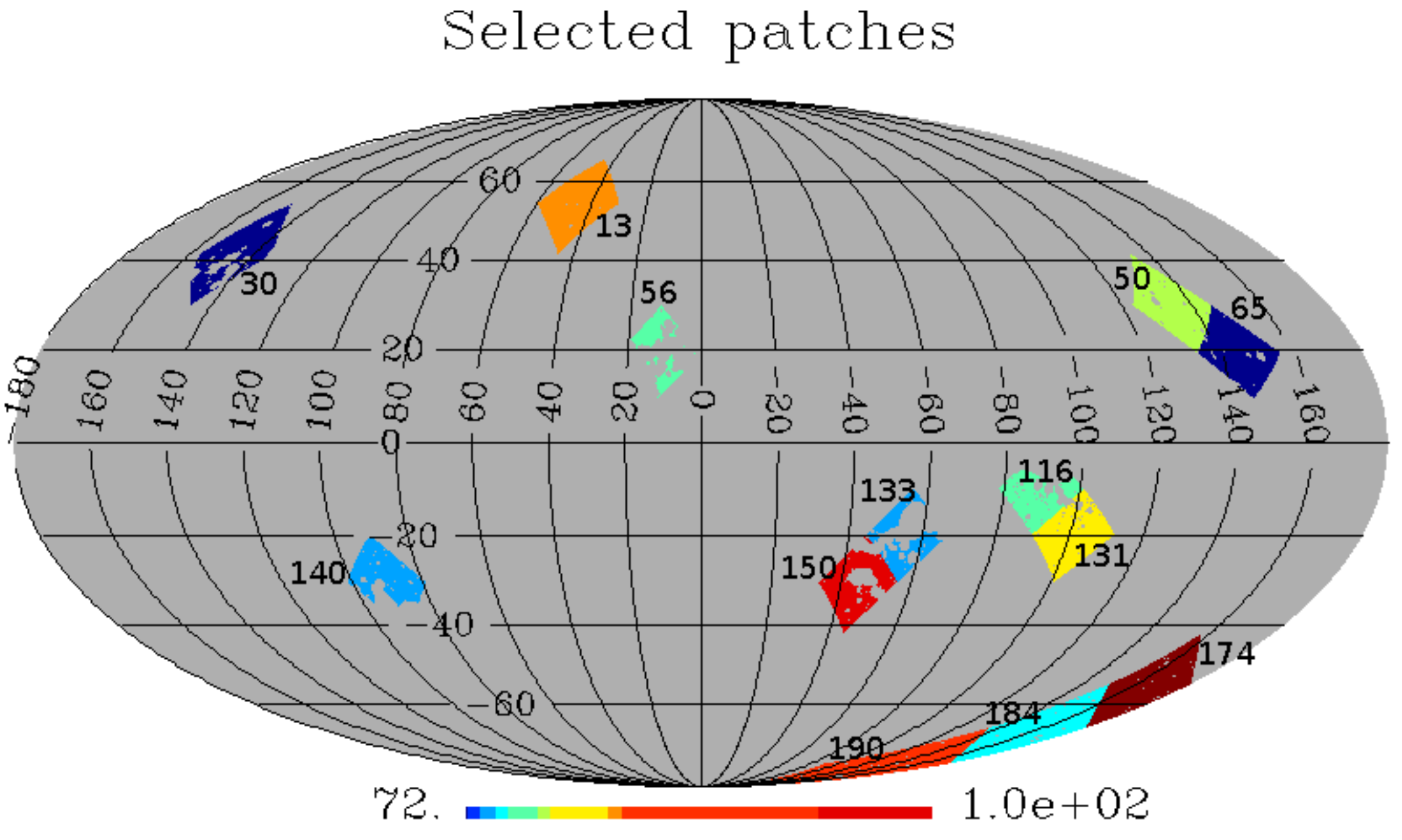}
\caption{Map highlighting the 13 regions selected according to the threshold $Total ~\chi^2 > 2.2 \sigma$ 
(see text for details).
The corresponding patch numbers are indicated. 
The color scale represents the total $\chi^2$ values of these regions, where the maximum value is 105 
corresponding to the patch \# 174. }
\label{fig5}
\end{figure}

\begin{figure*} 
\centering
\includegraphics[width=1\textwidth]{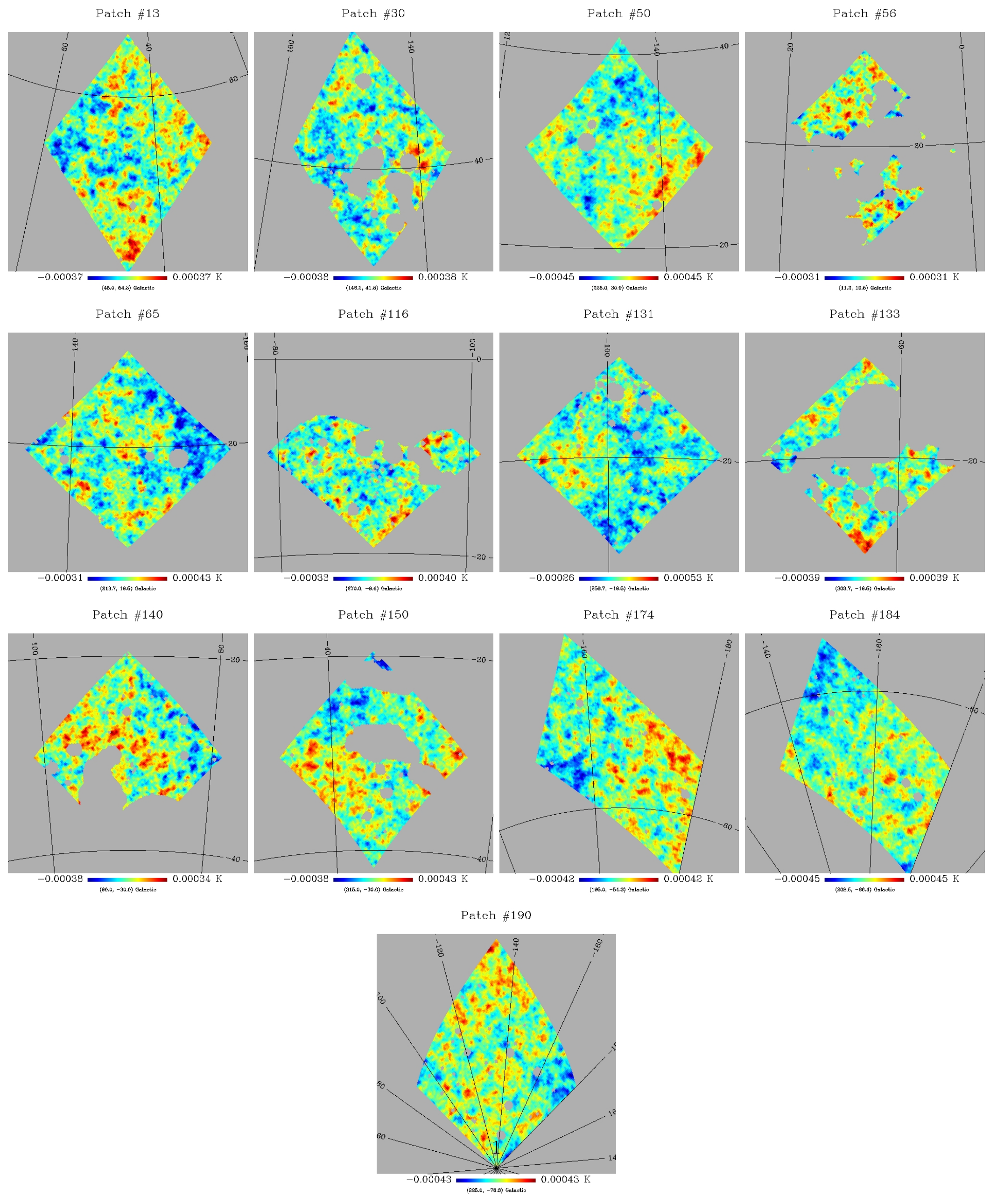}
\caption{For illustration we show the gnomview projections around the 13 anomalous regions 
from the $\mathtt{SMICA}$ CMB map .}
\label{fig6}
\end{figure*}

\noindent
Figure~\ref{fig3} shows the $total \, \chi^2$-map, while the Figure~\ref{fig4} graphically presents these total 
$\chi^2$ values as a function of the patch number. 
The horizontal dashed line represents the threshold we used as criterion for selecting the patches. 
This threshold value corresponds to the $2.2 \sigma$ deviation (dashed line) calculated upon the values of 
the $total \, \chi^2$-map. 
Finally, we selected all the patches whose total $\chi^2$ values are higher than $2.2 \sigma$ level.
As a result we find 13 sky patches showed in Figure~\ref{fig5}, in a full-sky Mollweide projection, 
and in Figure~\ref{fig6}, as individual gnomview projections, which exhibit NG at a high confidence 
level ($> 2.2 \sigma$).
Once these regions are selected, our MF analyses consider such 13 patches in the four foreground-cleaned 
Planck maps. 
Hereafter we call these patches \textit{anomalous regions}.

Aiming to investigate particular features, like the amplitude and signature, of the Gaussian deviations 
present in the selected anomalous regions, we compare their MF vectors, $\mathrm{v}_k$ 
(Equation \ref{def_v}), to the mean calculated from the MC Gaussian CMB maps. 
For this we use the \textit{relative difference} between them, 
defined as being the dif\/ference of both quantities normalized by the maximum value of the mean MF 
vector in analysis, that is, 
\begin{eqnarray}\label{eq-RD} 
\mbox{{\it relative difference}-MF}_k  \, \equiv \,
\frac{ \mathrm{v}_k^{Planck} - \langle \mathrm{v}_k \rangle }{\langle \mathrm{v}_k \rangle^{\mbox{\tiny MAX}}} \, , 
\end{eqnarray}
for $k = 1, 2$, i.e., Perimeter and Genus, respectively.

\begin{figure*} 
\centering
\includegraphics[width=0.8\textwidth]{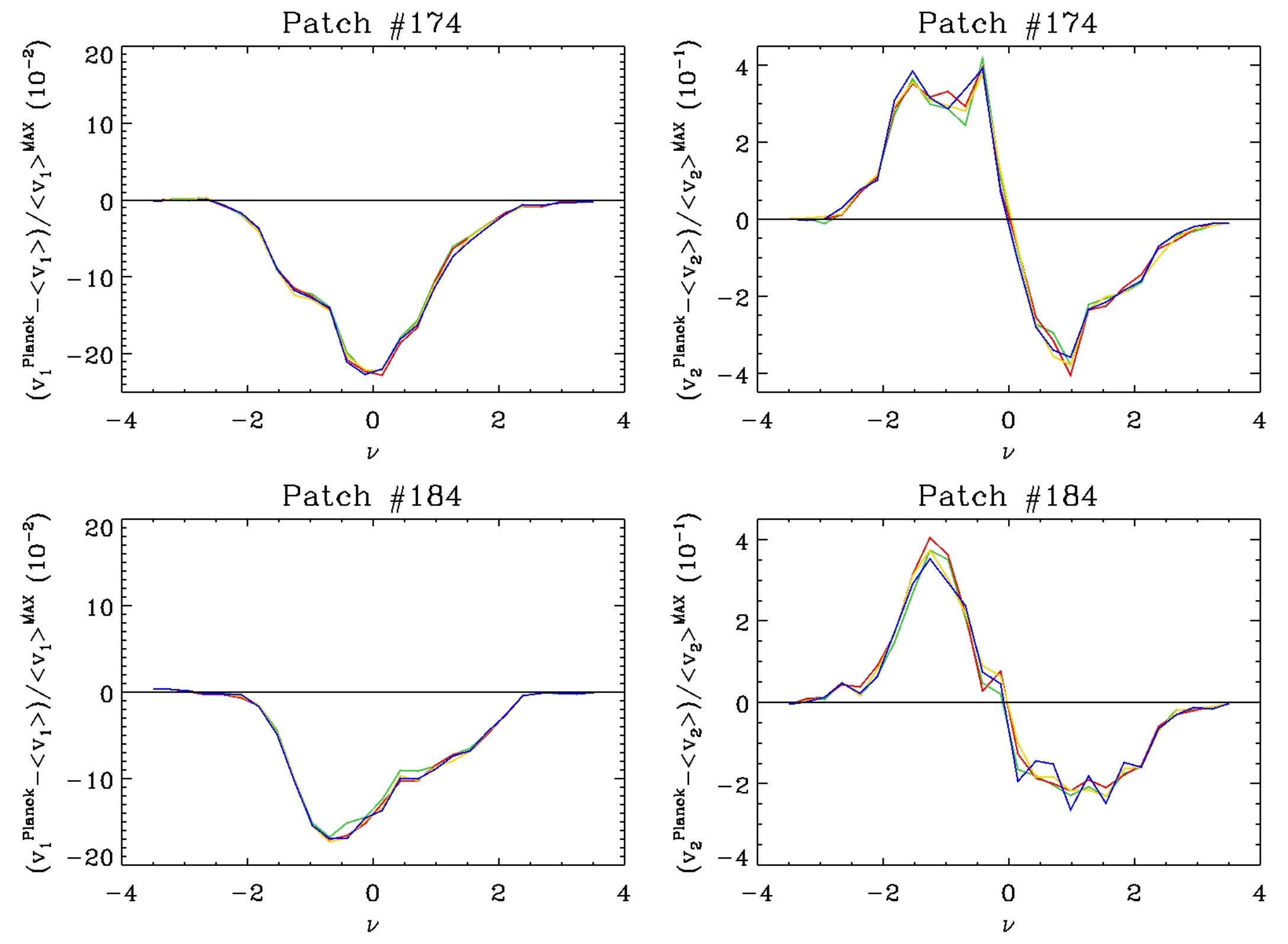}
\caption{Analyses of the anomalous regions encompassing the Cold Spot region. 
The plots displayed in the columns, left and right, correspond to the {\it relative difference}-MF for the 
Perimeter and Genus ($\mathrm{v}_k$, with $k = 1, 2$), respectively. 
The {\it relative difference}-MF was calculated for each foreground-cleaned Planck map ($\mathtt{SMICA}$, 
$\mathtt{NILC}$, $\mathtt{SEVEM}$ and $\mathtt{Commander}$ maps, corresponding to 
the blue, yellow, red, and green curves, respectively) according to the equation~\ref{eq-RD}, 
and the mean MF vectors ($\langle \mathrm{v}_k \rangle$) were obtained from the MC Gaussian CMB maps. 
All the analyses were performed using the $\mathtt{UT78}$ mask.}
\label{fig7}
\end{figure*}

\begin{figure*} 
\centering
\includegraphics[width=0.8\textwidth]{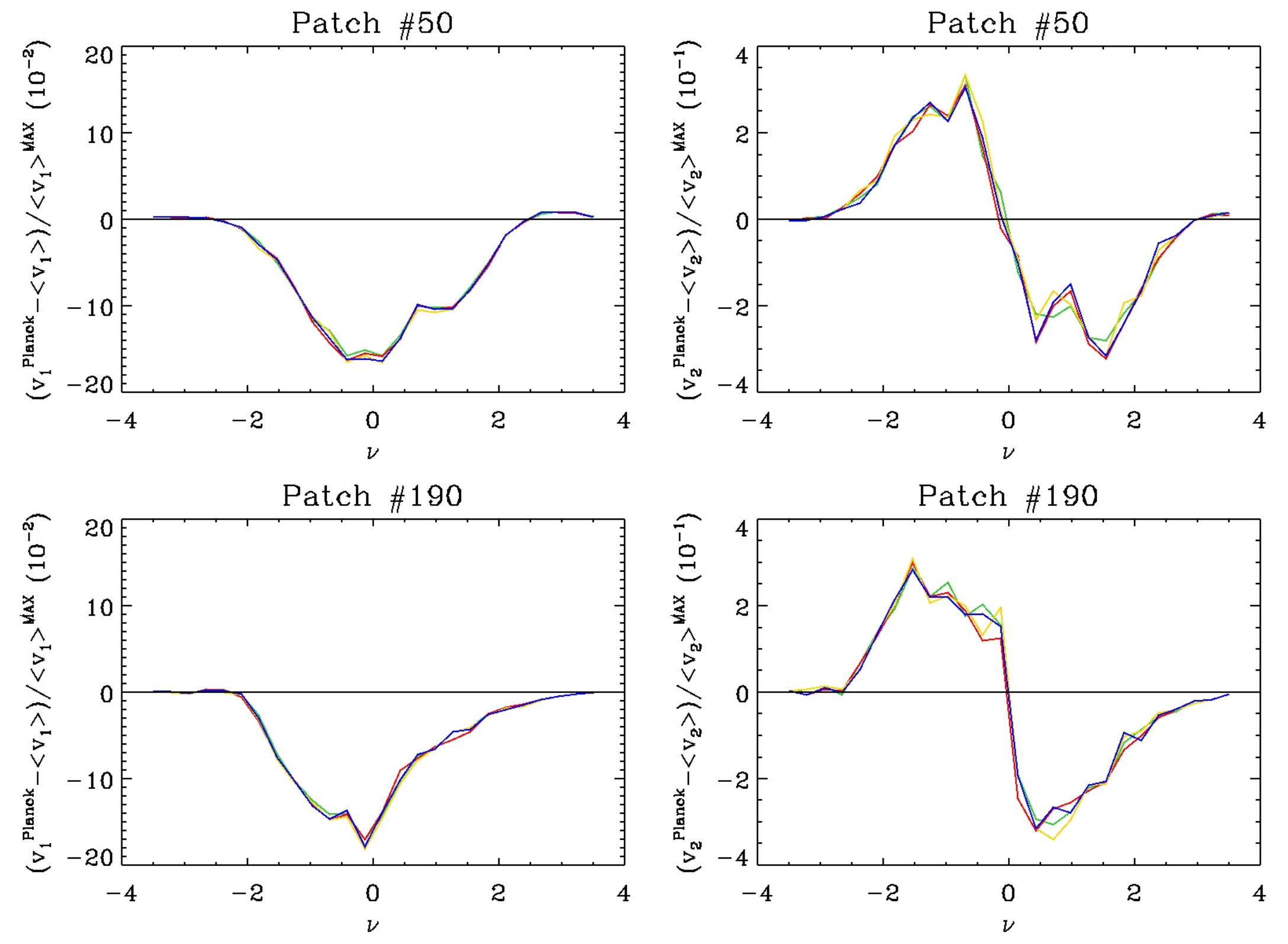}
\caption{Illustrative examples of the analyses of two patches, far from the galactic plane. 
The {\it relative difference}-MF curves reveal the contribution of a hot region in the patch \# 50 (positive $\nu$) and two cold regions in the patch \# 190 (negative $\nu$), as identified by the Planck collaboration 
(see the text for details).} 
\label{fig8}
\end{figure*}

\begin{figure*} 
\centering
\includegraphics[width=0.8\textwidth]{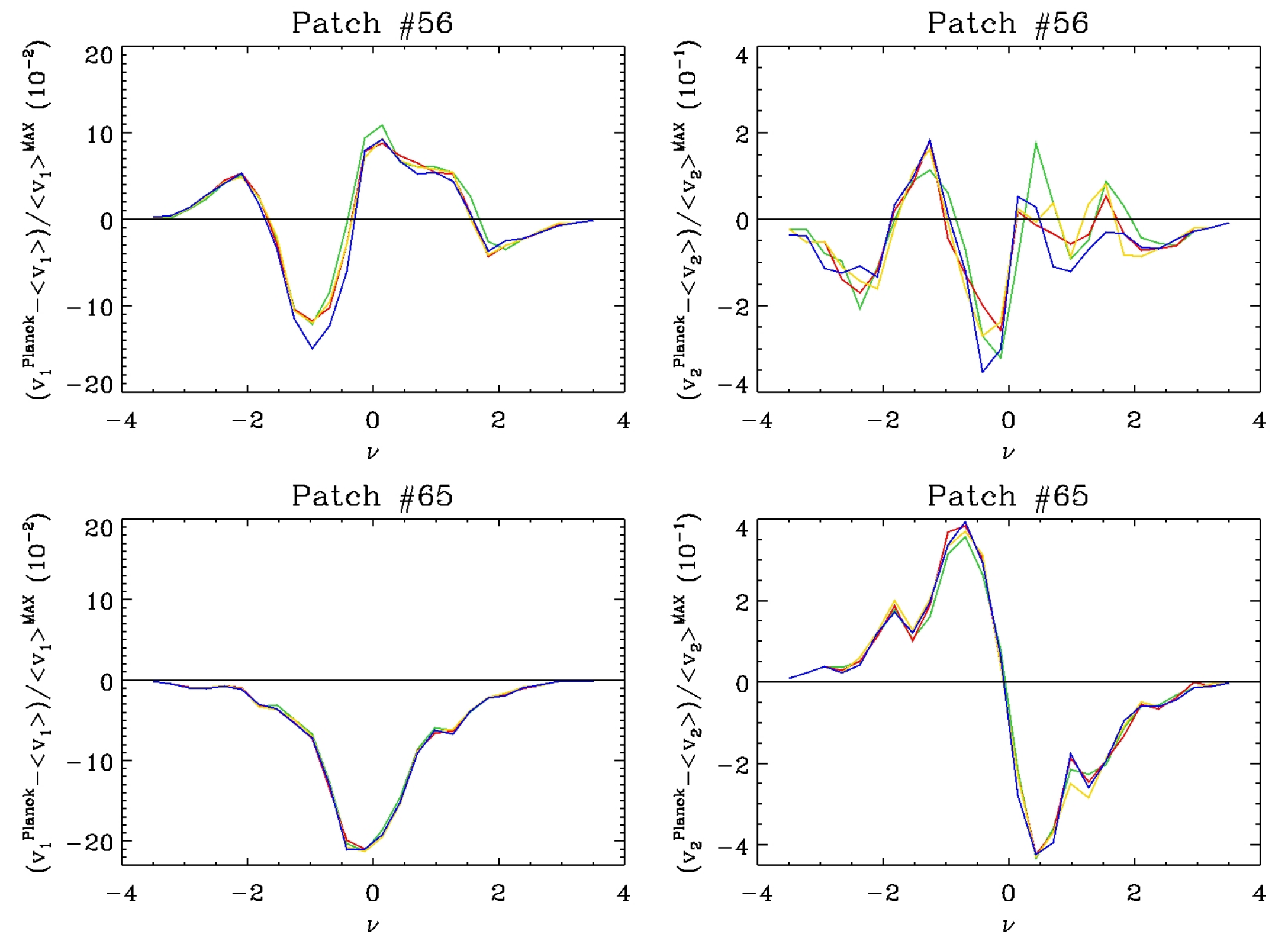}\\
\caption{Illustrative examples of the analyses of two patches, near the galactic plane. 
The {\it relative difference}-MF data for the patch \# 56 evidence noticeable dispersed curves 
(each color corresponds to each Planck map) suggestive of different residual foreground contaminations 
in each foreground-cleaned map; 
while the lower panels, for the patch \# 65, show the concomitant contribution from both a hot, previously 
found by the Planck collaboration, and a cold areas (see the text for details).}
\label{fig9}
\end{figure*}

The analyses of the four Planck CMB maps are presented in Figures \ref{fig7}-\ref{fig9}, that show 
illustrative examples of the resulting \textit{relative difference} curves for each MF of the anomalous 
regions (hereafter we call the curves displayed in Figures \ref{fig7}-\ref{fig9} and in the Appendix 
\ref{sec:appendix}, 
as the {\it relative difference}-MF). 
We divided the 13 anomalous regions in three sets, namely: 
\begin{itemize}
\item[I.] the regions near the Cold Spot~\citep{Vielva04}: patches \# 174 and 184 (see Figure \ref{fig7}) 
\item[II.] the regions far from the galactic plane: patches \# 13, 30, 50, 131, 140, 150, and 190 
(see Figure \ref{fig8})
\item[III.] the regions near to the galactic plane: patches \# 56, 65, 116, and 133 (see Figure 
\ref{fig9})
\end{itemize}

%
Before initiating the careful analyses of the {\it relative difference}-MF, we find important to verify 
the signature expected from such analyses.
For this, we investigate the presence of foreground residuals, 
inhomogeneous noise, and the effect of smoothing due to the application of a Gaussian beam to 
a CMB map, as discussed in Appendix \ref{sec:appendix}. 
The {\it relative difference}-MF resulting from such analyses show characteristic features 
when comparing two sets of Gaussian realizations with and without been smoothed by a Gaussian beam, 
which, except for their absolute amplitude, are very similar to those obtained when some foreground 
contamination is added (see Figure \ref{figA1}). 
In fact, one can observe the same general behavior from most of the curves presented in 
Figures~\ref{fig7}-\ref{fig9}, that is, a larger absolute amplitude around $\nu \sim 0$ for 
Perimeter, while in the Genus case this larger amplitude is usually observed for all $\nu$ values. 
Since the MF are quite sensitive to the smoothing scale of the temperature map \citep{2006/hikage}, 
we can infer that such pattern stems from the difference in the smoothing scales of synthetic 
and Planck data maps. 
Indeed, the effect of the beam in Planck data (\textsc{fwhm} = $5\arcmin$,~\citet{PLA2-IX}) makes 
the amplitude of the {\it relative difference}-MF systematically smaller than those obtained in Gaussian 
CMB synthetic maps, without a beam (compare Figures \ref{fig7}-\ref{fig9} with panels in the first row 
of Figure~\ref{figA1}). 
In addition, notice also that even with a reasonable concordance between the amplitude of the {\it relative 
difference}-MF obtained from the analysis of the Planck maps and the two last rows in Figure \ref{figA1}, 
for some of the selected patches such agreement occurs near the limit of the 2$\sigma$ regions, 
or still extrapolate it. 
Such behavior can be associated to differences between the effective beam of Planck maps 
and the Gaussian beam (\textsc{fwhm} = $5\arcmin$) we have used here, besides the addition of residual 
foreground contamination and inhomogeneous noise, which, as can be seen in Figure \ref{figA1}, 
increase the absolute amplitude of the {\it relative difference}-MF. 

It is interesting to compare our results with those obtained by the~\cite{PLA2-XVI}. 
In such analyses the authors used wavelets statistics, specifically the Spherical Mexican Hat wavelet 
(SMHW), to analyse Planck CMB maps finding a set of anomalous areas. 
The authors show that these regions correspond to cold and hot spots, corresponding to values 
below and above a given threshold of the SMHW coefficients, as a function of the wavelet scale. 
From this analysis they found 13 anomalous spots, of different sizes, considering several wavelet scales 
(\cite{PLA2-XVI}).
Comparing the sky position of each of these areas (see Figure 41 of \cite{PLA2-XVI}) with our 13 anomalous 
regions (Figure \ref{fig5}), we observe that 4 Planck areas do not have a correspondence within 
our analyses. 
Among the Planck anomalous spots with a counterpart in our analyses, those characterized by 
Planck hot areas are encompassed, totally or partially, by our patches 
\# 50, 131, 150, and 174,
while the Planck cold spots are related to our patches 
\# 65, 140, 174, 184, and 190. 
Notice that, there is no one-to-one correspondence between the Planck spots and ours patches, 
since some of our patches include more than one Planck's anomalous spot, as exemplified by our 
patch \# 174 which contains both a hot and a cold Planck spots. 
Generally speaking, although we miss 4 Planck areas, one can say that our results corroborate those 
obtained by the~\cite{PLA2-XVI}. 
Besides to confirm Planck anomalous spots, our scrutiny find 5 regions that also deserve special 
attention, namely, the patches 
\# 13, 30, 56, 116, and 133.

A distinctive feature observed in Figures \ref{fig7}-\ref{fig9} is the non-smooth behavior of the 
{\it relative difference}-MF curves as compared to the corresponding curves displayed in the 
Figure~\ref{figA1}, which appear smoothed because they correspond to the mean 
{\it relative difference}-MF obtained from a large set of MC realizations. 
Another noticeable feature observed in the analyses of simulated data (see Appendix A) concerns the 
symmetry property, in the Perimeter case, or anti-parity, in the Genus case, with respect to the 
$\nu = 0$ value.

Aware of such details, we start the scrutiny of the selected patches studying the signature produced by 
the Cold Spot (see, e.g.,~\cite{Vielva04, bernui09, Vielva10, zhao2014, PLA2-XVI} and references therein). 
As can be observed in Figure \ref{fig6}, this anomalous spot is encompassed by our patches 
\# 174 and 184. 
Its characteristic signature is clearly revealed by the Perimeter and Genus {\it relative-difference}-MF, 
as showed in the Figure \ref{fig7}. 
For these two patches, but specially for patch \# 184, it is possible to see the high amplitude of these 
curves skewed to negative $\nu$ values, suggestive of the presence of an anomalous cold region. 
The Cold Spot is the best-known CMB anomaly whose Perimeter {\it relative difference}-MF curve evidence 
a clear signature skewed to the left, that is, negative values of $\nu$, as noticed in Figure~\ref{fig7}.

Our patches \# 65, 140, and 190 include cold areas, as observed in Figure~\ref{fig6} and corroborated 
by the {\it relative difference}-MF curves in Figures \ref{fig8} and~\ref{fig9}, 
also detected by the Planck collaboration as being anomalous cold spots (although with diverse 
significance levels). 
In fact, the {\it relative difference}-MF curves from these regions confirm these facts 
(for illustrative purposes we displayed the case of the patch \# 190 in Figure~\ref{fig8} and 
the case of patch \# 65 in Figure~\ref{fig9}), but each one with its own features for $\nu < 0$, 
suggestive of their lower significance relative to the anomalous Cold Spot. 
Specifically, in region \# 65 we observe around $\nu = -2$ a small contribution from a cold area.
From this patch, the {\it relative difference}-MF, from both Perimeter and Genus, also show a contribution of 
what seems to be a hot area, around $1.0 < \nu < 1.5$, but not identified by Planck collaboration. 
As a matter of fact, the proximity of patch \# 65 to the Galactic plane is suggestive that the non-Gaussian 
contribution revealed in our analyses could be a mixture of residual unremoved foregrounds.

Moreover, one can also compare features revealed by the {\it relative difference}-MF, produced by 
anomalies of diverse significances associated to hot and cold areas. 
For this we analyze the results obtained for the patches 
\# 50, 131, 150, and 190, 
the first three corresponding to three hot areas, while the last one encompasses two cold areas, found by the \cite{PLA2-XVI}. 
Among these anomalies, the one claimed there as being the most significant is the hot area encompassed 
by the patch \# 150, whose hot nature is evident from Figure \ref{fig6}.
Its contribution can also be seen from the {\it relative difference}-MF, which evidence a deviation of its 
highest absolute amplitude for positive $\nu$ values.
Moreover, as we observe in Figure \ref{fig6}, besides the hot spot identified by the Planck collaboration, 
this patch also encompasses a cold spot which also leaves a clear signature upon negative $\nu$ values. 
As well as in the case of patch \# 150, the curves from \# 131, besides confirming the presence of the hot area identified by the Planck Collaboration, also suggest the presence of a small cold area, whose contribution occurs around $\nu = -1.5$ and can also be noted in Figure \ref{fig6}. 
Analyses of the {\it relative difference}-MF indicate that the second more significant hot area is encompassed 
by patch \# 50, showing its contribution for $1.0 < \nu < 1.5$, while \# 190 includes a large fraction of two 
cold areas, observed in the range $-2.0 < \nu < -0.5$ (see also Figure \ref{fig6}).

In agreement with the results of the \cite{PLA2-XVI}, where they show another anomaly reported as 
being a small hot area, we observe that it is also encompassed by patch \# 174. 
As a result of the concomitant presence of the Cold Spot, the {\it relative difference}-MF 
curves from patch \# 174 show 
a clear contribution of a cold area in the range $-2 < \nu < -1$ for both MF, besides a signal not so strong from a hot area, also evident in Figure \ref{fig6}. 
One can see a slight contribution of this hot anomaly around $\nu \sim 1$ in the Genus curve, but it is not 
clear from the Perimeter. 
This feature suggests that the presence of only a small fraction of the Cold Spot inside the patch \# 174, 
is still able to manifest a dominant signature compared with the one due to the hot spot. 
This can be an indication of why some anomalies found by the \cite{PLA2-XVI} do not have a 
corresponding finding within our analyses, that is, their signals appear diluted in the regions considered 
here. 
In this scenario, one understands the necessity of using different estimators: 
since one does not expect any single estimator to be able to recognize the type, amplitude, and origin of 
all the signals contributing to some detected Gaussian deviation signal, it is of utmost importance to use a 
variety of complementary tools to reveal them.

Moreover, in addition to confirm the Planck findings, our analyses also aimed the identification of other 
anomalous regions. 
The regions whose corresponding {\it relative difference}-MF show features suggestive of the presence of 
significant cold areas are \# 13, 30, and 56. 
The examination of the {\it relative difference}-MF of other two anomalous patches, namely, 
\# 116 and 133, both near the galactic region, reveals some information. 
For the patch \# 133, one observes, in special for the Perimeter case, a high amplitude of the curve skewed 
for positive $\nu$ values, indicative that it is sourced by a hot spot, which, due to its proximity to the 
galactic plane, could be a residual foreground. 
In the case of patch \# 116, the {\it relative difference}-MF curves suggest the presence of both cold 
and hot spots, although their presence are not visually revealed in Figure \ref{fig6}. 
Additionally, it is interesting to mention that the {\it relative difference}-MF for patches \# 56 and 116
exhibit an unexpected behavior as compared with Gaussian CMB patches (see, for comparison, 
Figure~\ref{fig9} and Figure~\ref{figA1}).

Finally, it is worth to stress the importance in testing the four foreground-cleaned Planck CMB maps because 
the applied procedures, although highly efficient, could have leaved undesired residuals. 
As a result of this, data maps are a mixture of the primordial temperature fluctuations and residual signals 
of secondary origin \citep{2015/axelsson}. 
Moreover, it is not expected that possible residual contaminations in each Planck CMB map would be 
of the same type and intensity, because each one was obtained with a different component separation 
procedure. 
This fact can explain the behavior presented by the {\it relative difference}-MF curves in some patches, 
for instance, in the patch \# 56 displayed in Figure \ref{fig9}, 
where it is observed a dispersion 
of the colored curves corresponding to the four Planck maps, suggestive of residual contaminations of 
diverse types and intensities.

\section{Conclusions and final Remarks} \label{sec:section6}

The search for non-Gaussian signals in CMB datasets has become the scientific target of several teams. 
Recent analyses from the~\cite{PLA2-XVI} reported 13 anomalous hot and cold spots, of diverse 
angular sizes in local analyses of the foreground-cleaned Planck CMB data. 

We have performed detailed analyses of the four foreground-cleaned Planck CMB maps aiming to identify 
non-Gaussian signals in sky patches and to describe their characteristic signatures. 
For this purpose we divided the sky in a total of 192 regions, and used the three MF to analyze each one. 
It is well-known that the MF calculated at a sequence of thresholds $\nu$, as well as the diverse MF, are correlated.  
For this, our joint estimator is defined in Equation (\ref{eq:chi2}) considering the full covariance 
matrix~\citep{2006/hikage,2012/ducout}. 
The methodology adopted in this work let us to find a set of 13 regions in the Planck CMB maps, which, 
according to the MF analyses, show large discrepancies with respect to what is expected in a Gaussian 
CMB temperature field. 
Such regions have large $\chi^2$ values in disagreement with the mean obtained from Gaussian 
realizations in more than $2.2 \sigma$, and, therefore, classified here as anomalous. 
Our scrutiny give special attention to the signature of the non-Gaussian deviations revealed by the MF 
applied to Planck data, in comparison with similar analyses done in MC Gaussian CMB maps. 
The obtained results lead us to the following conclusions: 

\begin{itemize}
\item 
The comparison between the $\chi^2$-maps obtained using the Area and those combining the 
other two MF to analyse the four Planck CMB maps (Figure~\ref{fig1}) confirms the lower sensitivity of 
this MF relatively to the other two (noticed before by~\cite{2012/ducout} and \cite{Novaes14,Novaes15}). 
\item 
The absence of a definite dependence among the $\chi^2$ values and the percentage of valid pixels, 
showed in Figure~\ref{fig2}, evidence a weak dependence of the patch effective area in our analyses. 
\item 
The dif\/ferent $\chi^2$ values obtained in the Planck CMB maps could be an indication of the presence 
of residual contaminations in these foreground-cleaned CMB maps. 
However, we observe that both $\chi^2$-maps, calculated for the 
four Planck maps, as well as the $total \,\chi^2$-map reveal the highest values for the same 13 anomalous 
patches, evidencing that the NG there have common features. 
\end{itemize}

As discussed above, our inferences are mainly based upon the estimated $\chi^2$ values, 
which locally compare the Planck maps to a set of MC Gaussian CMB maps. 
In this sense, we have used the total $\chi^2$ values, as represented in Figure \ref{fig4}, to find 13 regions 
with the largest discrepancy from Gaussianity, that is, they are the most anomalous regions. 
A direct comparison of the sky position of our selected patches to the results presented in Figure 41 
of the \cite{PLA2-XVI} allows us to affirm that our results not only corroborate Planck findings, but also 
reveal possibly new anomalous regions. 

As important as the selection of such anomalous regions is the careful analysis of the signature 
produced by them and revealed by the MF. 
Once more we confirm statements from the \cite{PLA2-XVI}, since the specific signatures of each 
region lead us to associate these anomalous regions to cold and hot areas. 
Additionally, we found 5 new regions, namely, patches \# 13, 30, 56, 116, and 133.
Ultimately, note that regions near the Galactic plane (e.g., patches \# 56, 116, and 133) are suspicious 
of under-subtraction (over-subtraction) of contaminations in the foreground cleaning process, leaving 
hot (cold) regions. 
In fact, we probably have a mixture of primordial signal and a small contribution of secondary residuals 
not only near the galactic region, but also in other regions of the CMB maps (see~\citet{Aluri12}), what 
makes the type of analysis we have performed here of utmost importance.

\section*{Acknowledgments}
\noindent
We acknowledge the use of the code for calculating the MF, from~\cite{2012/ducout} and~\cite{2012/gay}. 
Some of the results in this paper have been derived using the HEALPix package~\citep{2005/gorski}. 
CPN and GAM acknowledge Capes fellowships. 
AB acknowledges f\/inancial support from the Capes Brazilian Agency, for the grant 88881.064966/2014-01. 
We thank Glenn Starkman for insightful comments and suggestions.


\appendix

\section{Signature from secondary sources} \label{sec:appendix}

Aiming to support some of our conclusions and propitiate the reader an easier way to verify our claims, 
we performed some additional tests evaluating how the presence of secondary signals and the beam effect 
would influence our analyses. 
For this we investigate the amplitude and signature of the {\it relative difference}-MF obtained comparing the 
ideal MC Gaussian CMB maps used throughout this paper and three other sets of synthetic CMB maps. 
These sets are composed by 5000 MC realization each, generated considering 
${\rm \,N}_{\mbox{\small side}} = 512$, and corresponds to Gaussian CMB maps to which we add the 
following effects:

\begin{itemize}
\item[(a)] smoothing these MC Gaussian CMB maps with a Gaussian beam with \textsc{fwhm} = 5\arcmin,
\item[(b)] adding a residual contribution of foreground emission (as explained below) 
and a Gaussian beam with \textsc{fwhm} = 5\arcmin,
\item[(c)] adding inhomogeneous noise (as explained below) 
and a Gaussian beam with \textsc{fwhm} = 5\arcmin.
\end{itemize}

The first set of maps, set (a), is simulated as described in Section \ref{sec:section4.1}, posteriorly smoothed 
with the HEALPix Gaussian beam of \textsc{fwhm} = 5\arcmin. 
The second one, set (b), is constructed including the contribution from foreground emission as a residual 
signal, in an attempt to imitate the possible contents of a Planck data map. 
Initially, we estimate the contribution of the most important Galactic foreground signals, namely, the 
synchrotron, free-free, and dust emission. 
For this we performed an extrapolation (or interpolation in the case of the dust emission), pixel by pixel, from 
a set of data maps covering a wide range of frequencies, maps publicly available as part of the Planck and 
WMAP-9yr data releases~\citep{PLA2-X, Bennett13wmap}. 
The residual contribution, estimated as being 10\% of the final template, was then added to the Gaussian 
MC CMB maps also, generated as described in Section \ref{sec:section4.1}.
Finally, we have applied a Gaussian beam of \textsc{fwhm} = 5\arcmin~upon these maps.

The simulation of the inhomogeneous noise considered in the set (c) was performed based on what is 
expected to be present in the $\mathtt{SMICA}$ map. 
For this we used the noise map released jointly to the $\mathtt{SMICA}$ map to estimate the $\sigma_{noise}$ 
in each pixel $p$ as being the standard deviation of the noise values from a set of pixels in a disk around $p$. 
Multiplying the standard deviation map, pixel-by-pixel, by a normal distribution with zero mean and unitary 
standard deviation we obtain the $\mathtt{SMICA}$-like noise map.
The noise map is added to the Gaussian realizations already convolved with a HEALPix Gaussian beam 
with \textsc{fwhm} = 5\arcmin. 

\begin{figure*}
\centering
\includegraphics[width=0.7\textwidth]{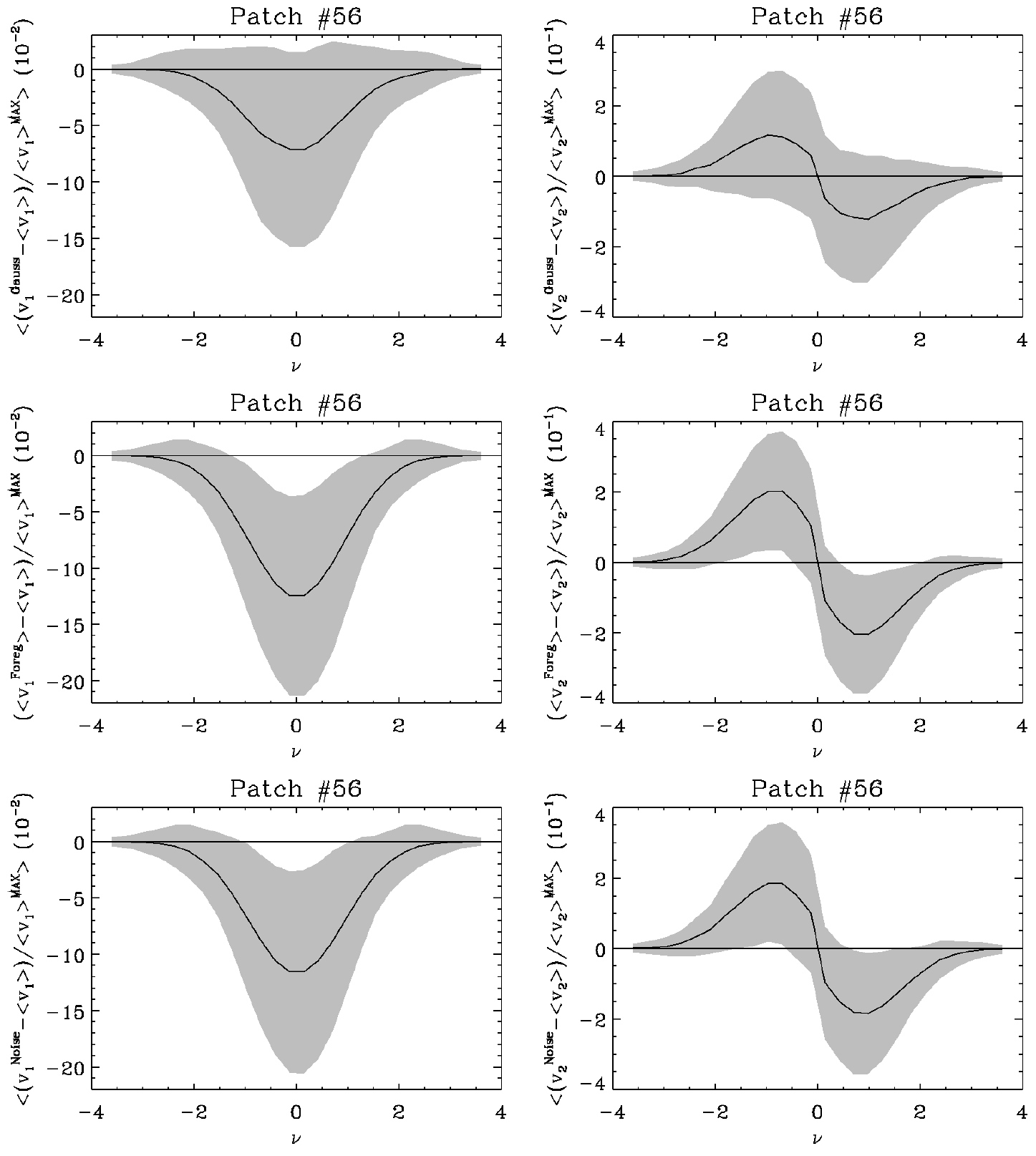}
\caption{Plots of the mean {\it relative difference}-MF obtained comparing the Perimeter (left) and 
Genus (right) vectors calculated, from top to bottom, for the datasets (a), (b), and (c), and the average 
obtained from the ideal MC Gaussian CMB maps used in the analyses presented throughout the paper. 
Such curves are obtained from the analyses done in the patch \# 56. 
The gray region represents the 2$\sigma$ level (see text for details).}
\label{figA1}
\end{figure*}

For each map of these three sets we calculate the Perimeter and Genus vectors of the sky region corresponding to the 
patch \# 56. 
This patch was chosen for two reasons, namely, its proximity to the Galactic plane and higher noise contribution. 
The Figure \ref{figA1} presents the mean {\it relative difference}-MF between the $k$-th MF obtained from each 
map of a given set of realizations and the mean MF calculated from the ideal MC Gaussian 
CMB maps used throughout the analyses of previous sections, $\langle v_k \rangle$, that is,

\begin{equation}
\langle \mbox{rel. difference-MF}_k^j \rangle \equiv
\bigg \langle \bigg ( \frac{\mathrm{v}_k^j  - \langle \mathrm{v}_k \rangle}{\langle \mathrm{v}_k \rangle^{\mbox{\tiny MAX}}} \bigg ) \bigg |_{\mbox{for the {\it i}-th map}} \bigg\rangle, 
\end{equation}
where $i = 1, ..., 5000$, and the index $j$ indicates the data-sets, that is, 
$j=$ (a), then $\mathrm{v}_k^j =  \mathrm{v}_k^{\mbox{\tiny Gauss.}}$, 
$j=$ (b), then $\mathrm{v}_k^j =  \mathrm{v}_k^{\mbox{\tiny Foreg.}}$, and 
$j=$ (c), then $\mathrm{v}_k^j =  \mathrm{v}_k^{\mbox{\tiny Noise}}$. 
Note from this equation that the mean curve is the average of the {\it relative difference}-MF obtained 
from the set of MC CMB maps. 
%


\bsp
\label{lastpage}

\end{document}